\documentclass{moriond}

\usepackage{subcaption}
\usepackage{aas_macros}

\def\be{\begin{equation}}
\def\ee{\end{equation}}
\def\bea{\begin{eqnarray}}
\def\eea{\end{eqnarray}}

\newcommand{\planck}{{\em Planck}}

\newcommand{\Photo}{}

\begin{document}
\vspace*{4cm}
\title{MASSIVE MERGING CLUSTER PSZ2G091 AS SEEN BY THE NIKA2 CAMERA}

\author{\scriptsize E. ~Artis$^{1}$,
    R. ~Adam$^{2}$,
     P. ~Ade$^{3}$, 
      H. ~Ajeddig$^{4}$, 
     P.  ~Andr\'e$^{4}$, 
    M. ~Arnaud$^{4}$, 
     H. ~Aussel$^{4}$, 
     I. ~Bartalucci$^{5}$, 
     A. ~Beelen$^{6}$, 
     A. ~Beno\^it$^{7}$, 
     S. ~Berta$^{8}$, 
     L. ~Bing$^{9}$, 
     O. ~Bourrion$^{1}$, 
     M. ~Calvo$^{7}$, 
     A. ~Catalano$^{1}$, 
     M. ~De~Petris$^{10}$,
     F.-X. ~D\'esert$^{11}$, 
     S. ~Doyle$^{3}$, 
     E.~F.~C. ~Driessen$^{8}$, 
     A. ~Ferragamo$^{10}$,
     A. ~Gomez$^{12}$, 
     J. ~Goupy$^{7}$,
     C. ~Hanser$^{1}$,
     F. ~K\'eruzor\'e$^{1}$, 
     C. ~Kramer$^{13}$, 
     B. ~Ladjelate$^{13}$, 
     G. ~Lagache$^{9}$, 
     S. ~Leclercq$^{8}$, 
     J.-F. ~Lestrade$^{14}$, 
     J.-F. ~Mac\'ias-P\'erez$^{1}$, 
     A. ~Maury$^{4}$, 
     P. ~Mauskopf$^{3,20}$, 
    F. ~Mayet$^{1}$, 
     A. ~Monfardini$^{7}$,
    M. ~Mu\~noz-Echeverr\'{\i}a$^{1}$,
     A. ~Paliwal$^{10}$, 
     L. ~Perotto$^{1}$,
     G. ~Pisano$^{3}$, 
     E. ~Pointecouteau$^{15}$, 
     N. ~Ponthieu$^{11}$, 
     G.~W. ~Pratt$^{4}$,
     V. ~Rev\'eret$^{4}$, 
     A.~J. ~Rigby$^{3}$, 
     A. ~Ritacco$^{16,6}$, 
     C. ~Romero$^{17}$, 
     H. ~Roussel$^{18}$, 
     F. ~Ruppin$^{19}$, 
     K. ~Schuster$^{8}$, 
     S. ~Shu$^{21}$,
     A. ~Sievers$^{13}$, 
     C. ~Tucker$^{3}$, 
     G. ~Yepes$^{22}$, 
}

\address{ \scriptsize $^1$Univ. Grenoble Alpes, CNRS, LPSC-IN2P3, 53, avenue
des Martyrs, 38000 Grenoble, France.\\
     $^2$LLR, CNRS, École Polytechnique, Institut Polytechnique de Paris,
Palaiseau, France.\\
     $^3$School of Physics and Astronomy, Cardiff University, Queen’s
Buildings, The Parade, Cardiff, CF24 3AA, UK.\\
     $^4$AIM, CEA, CNRS, Universit\'e Paris-Saclay, Universit\'e Paris
Diderot, Sorbonne Paris Cit\'e, 91191 Gif-sur-Yvette, France.\\
     $^5$INAF, IASF-Milano, Via A. Corti 12, 20133 Milano, Italy.\\
     $^6$Institut d'Astrophysique Spatiale (IAS), CNRS, Universit\'e Paris
Sud, Orsay, France.\\
     $^7$Institut N\'eel, CNRS, Universit\'e Grenoble Alpes, France.\\
     $^8$Institut de RadioAstronomie Millim\'etrique (IRAM), Grenoble, France.\\
     $^9$Aix Marseille Univ, CNRS, CNES, LAM, Marseille, France.\\
     $^{10}$Dipartimento di Fisica, Sapienza Universit\`a di Roma, Piazzale
Aldo Moro 5, I-00185 Roma, Italy.\\
     $^{11}$Univ. Grenoble Alpes, CNRS, IPAG, 38000 Grenoble, France.\\
     $^{12}$Centro de Astrobiolog\'ia (CSIC-INTA), Torrej\'on de Ardoz, 28850
Madrid, Spain.\\
     $^{13}$Instituto de Radioastronom\'ia Milim\'etrica (IRAM), Granada, Spain.\\
     $^{14}$LERMA, Observatoire de Paris, PSL Research University, CNRS,
Sorbonne Universit\'e, UPMC, 75014 Paris, France.\\
     $^{15}$Univ. de Toulouse, UPS-OMP, CNRS, IRAP, 31028 Toulouse, France.\\
     $^{16}$Laboratoire de Physique de l’\'Ecole Normale Sup\'erieure, ENS, PSL
Research University, CNRS, Sorbonne Universit\'e, Universit\'e de Paris,
75005 Paris, France.\\
     $^{17}$Department of Physics and Astronomy, University of Pennsylvania,
209 South 33rd Street, Philadelphia, PA, 19104, USA.\\
     $^{18}$Institut d'Astrophysique de Paris, CNRS (UMR7095), 98 bis boulevard
Arago, 75014 Paris, France.\\
     $^{19}$Univ. Lyon, Univ. Claude Bernard Lyon 1, CNRS/IN2P3, IP2I Lyon, F‐69622, Villeurbanne, France.\\
     $^{20}$School of Earth and Space Exploration and Department of Physics,
Arizona State University, Tempe, AZ 85287, USA.\\
     $^{21}$Caltech, Pasadena, CA 91125, USA.\\
     $^{22}$Departamento de F\'isica Te\'orica and CIAFF, Facultad de Ciencias,
Modulo 8, Universidad Aut\'anoma de Madrid, 28049 Madrid, Spain.}

\maketitle\abstracts{
  PSZ2 G091.83+26.11 is a galaxy cluster with $M_\mathrm{500}= 7.43\times10^{14} M_\odot$ at $ z = 0.822$ \cite{2016A&A...594A..27P}. This object exhibits a complex morphology with a clear bimodality observed in X-rays. However, it was detected and analysed in the \planck~sample as a single, spherical cluster following a universal profile \cite{2010A&A...517A..92A}. This model can lead to miscalculations of thermodynamical quantities, like the pressure profile. As future multiwavelength cluster experiments will detect more and more objects at high redshifts, it is crucial to quantify this systematic effect.
In this work, we use high-resolution observations of the NIKA2 camera \cite{2020A&A...637A..71P}$^,$\cite{2018A&A...609A.115A}$^,$\cite{2016JLTP..184..816C}$^,$\cite{2016JInst..1111101B} to integrate the morphological characteristics of the cluster in our modelling. This is achieved by fitting a two-halo model to 
the SZ image and then by reconstruction of the resulting projected pressure profile. We then compare these results with the spherical assumption.}
\section{Cosmological inference from cluster counts}
\label{intro}
The number of clusters per unit of mass and volume, modelled as the halo mass function \cite{1974ApJ...187..425P}, constitutes a robust cosmological probe \cite{2011ARA&A..49..409A}. However, the total mass of dark matter halos is not an observable quantity, and must be inferred from different physical phenomena \cite{2019SSRv..215...25P}, like the thermal Sunyaev-Zel’dovich effect \cite{1970Ap&SS...7....3S} (tSZ). As a consequence, astrophysical systematic effects are biasing our measurements \cite{Bocquet2015}, and must be integrated in any cosmological analysis. Moreover, given the fact that the size of the catalogs will be increased by several orders of magnitude in the future, effects that are now neglected will play a crucial role \cite{2021A&A...649A..47A}$^,$\cite{2021A&A...652A..21F}. \\
The Large program SZ \cite{2020EPJWC.22800017M} of the NIKA2 experiment (LPSZ) aims at investigating these issues, taking advantage of the spatial resolution and FoV of the NIKA2 camera.  Here, we focus on the impact of cluster morphology on the reconstruction of thermdynamical quantities.
\section{The case of PSZ2G091}
\label{sec:psz2G091}
\begin{figure}[h]
\centering
\begin{minipage}{.33\textwidth}
  \centering
  \includegraphics[width=\linewidth]{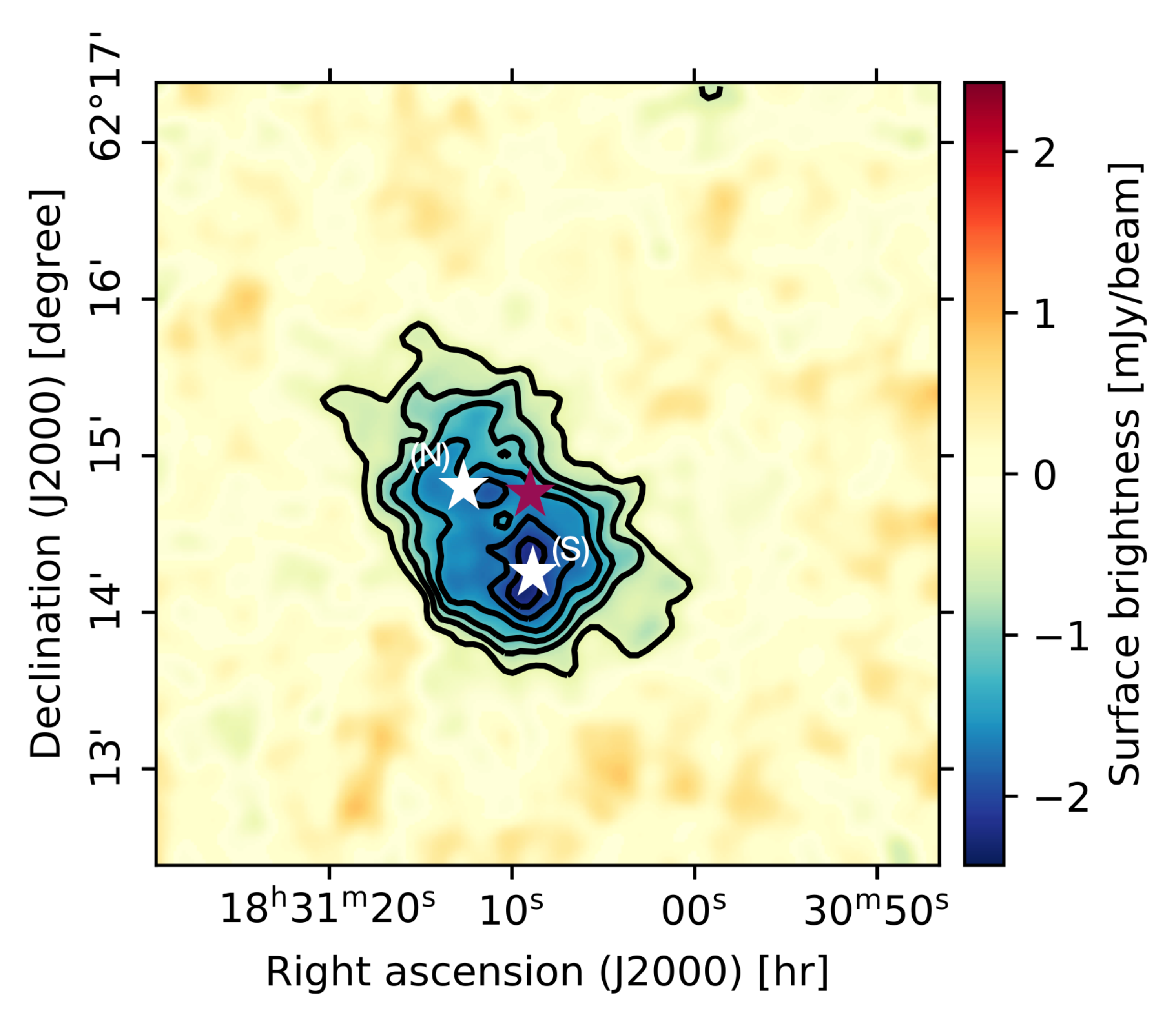}
\end{minipage}%
\begin{minipage}{.33\textwidth}
  \centering
  \includegraphics[width=\linewidth]{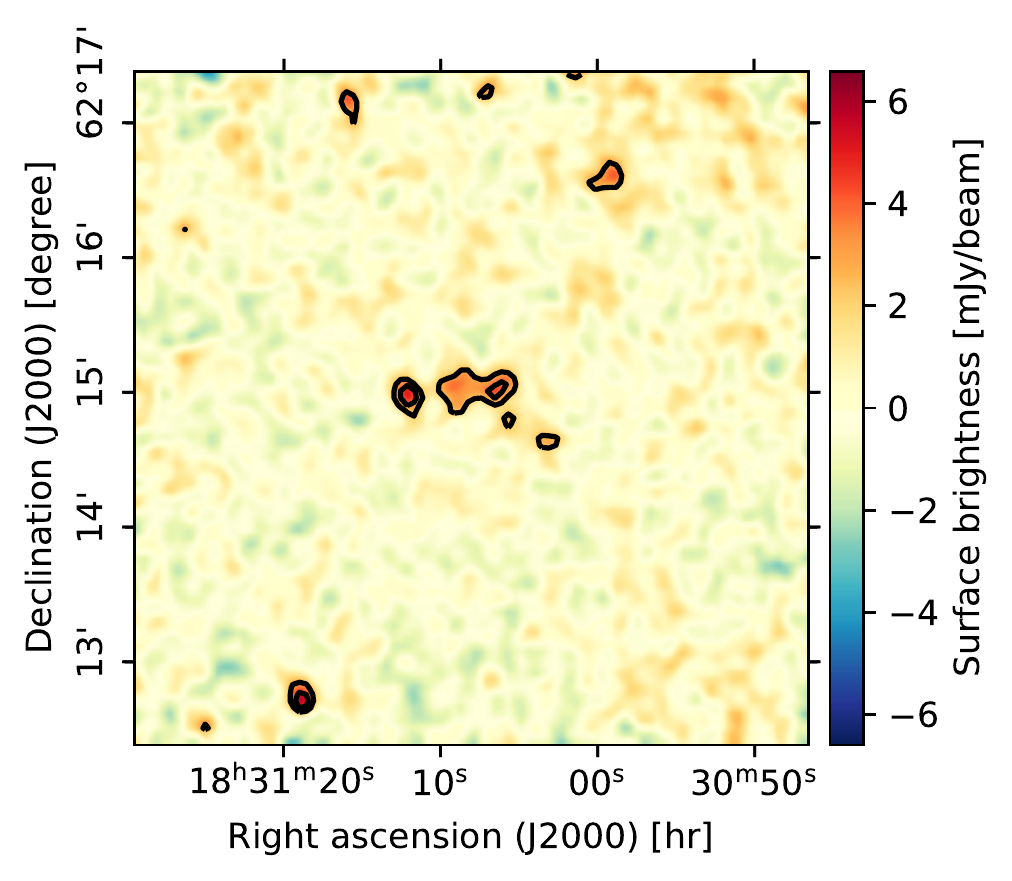}
\end{minipage}
\begin{minipage}{.33\textwidth}
  \centering
  \includegraphics[height=4.7cm,width=\linewidth]{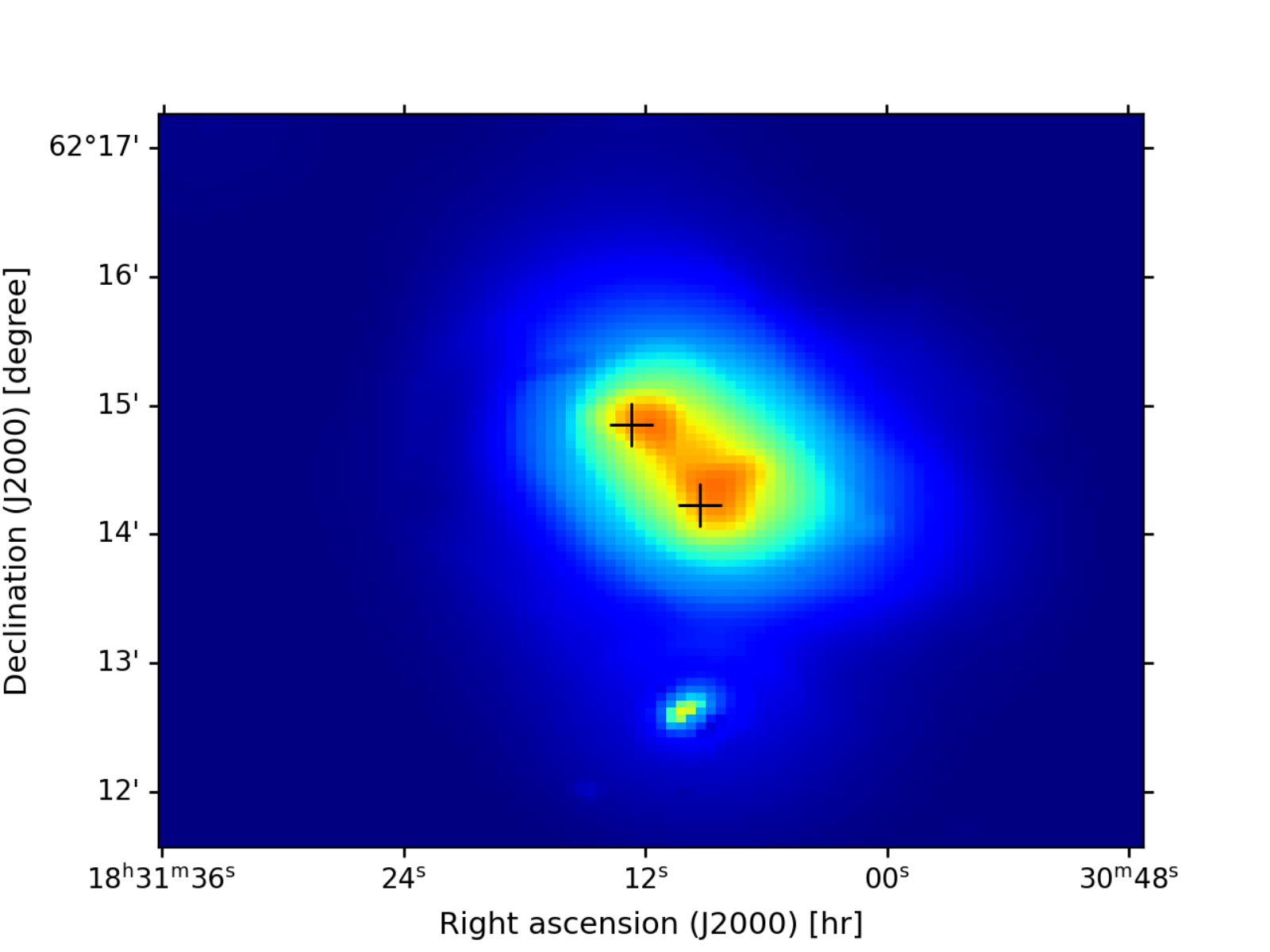}
\end{minipage}

\caption{\footnotesize \textbf{Left}: Surface brightness map at 2 mm from NIKA2 observations. The black contours represent the SNR levels starting at  $3\sigma$ increasing by $2\sigma$ at each step. The northern (N) and southern (S) X-ray peaks are shown as white stars, the X-ray centroid being in purple. \textbf{Middle}: Surface brightness at 1 mm. The signal exhibits several point sources, which are fitted in the analysis. \textbf{Right}: X-ray surface brightness of PSZ2G091 from \textit{XMM-Newton}. The cluster exhibits two peaks at the position $(\alpha_\mathrm{N},\delta_\mathrm{N})=(277.80,62.24)$, and $(\alpha_\mathrm{S},\delta_\mathrm{S})=(277.79, 62.237)$, shown with the black crosses.}
\label{fig:nika2_2mm}
\end{figure}
\begin{table}[t]
\caption[]{\footnotesize Characteristics of PSZ2 G091.83+26.11  \cite{2016A&A...594A..27P}. $t_\mathrm{LPSZ}$ is the scheduled duration of the observations, while  $t_\mathrm{obs}$ is the actual time during which the object has been observed. }
\label{tab:cluster_car}
\vspace{0.4cm}
\begin{center}
\begin{tabular}{|c|c|c|c|c|}
\hline
 $z$ & $M_\mathrm{500}$ & $\theta_\mathrm{500}$ & $t_\mathrm{obs}/t_\mathrm{LPSZ}$& tSZ decrement peak  \\ \hline 
0.822 & $7.43\times 10^{14}M_\odot$ & $2.2~\mathrm{arcmin}$ & 2.5h/2.5h=1 & 14.9$\sigma$ \\ \hline
\end{tabular}
\end{center}
\end{table}
\noindent As part of the LPSZ, PSZ2G091 was observed in October, 2017, with an average elevation of $58.5^\circ$ and an average atmospheric opacity at 225 GHz of 0.243. These conditions are standard for observations at the IRAM 30 m telescope, at this season.\\
In figure \ref{fig:nika2_2mm}, we show the results of the data reduction  \cite{2020A&A...637A..71P} at 1 and 2 mm. The cluster is clearly elongated in the NE-SW direction. There is a clear departure from sphericity, and a hint of bimodality later confirmed in the X-ray surface brightness map. The peaks in the X-ray map are in good agreement with the ones observed in the NIKA2 2 mm map. This would imply the presence of two well-defined sub-halos in the first stages of a major merger.
\section{Imaging analysis}

\begin{figure}
\centering
\begin{subfigure}[b]{0.6\textwidth}
   \includegraphics[width=1\linewidth]{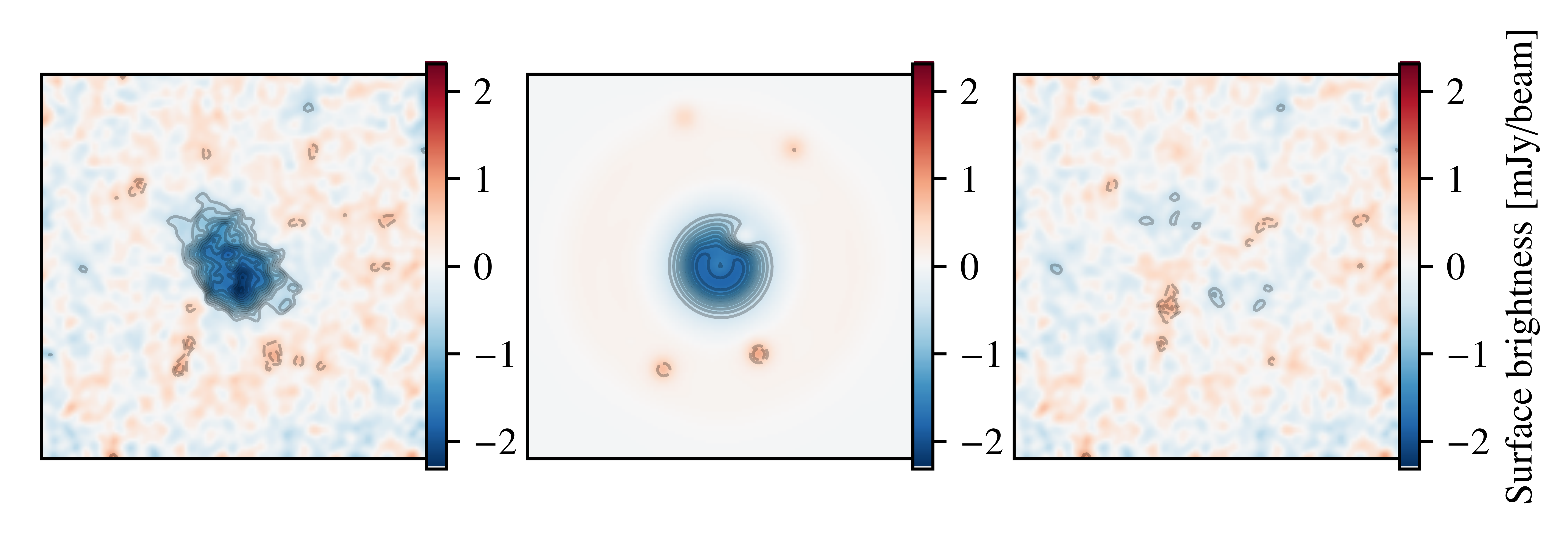}
   \end{subfigure}

\begin{subfigure}[b]{0.6\textwidth}
   \includegraphics[width=1\linewidth]{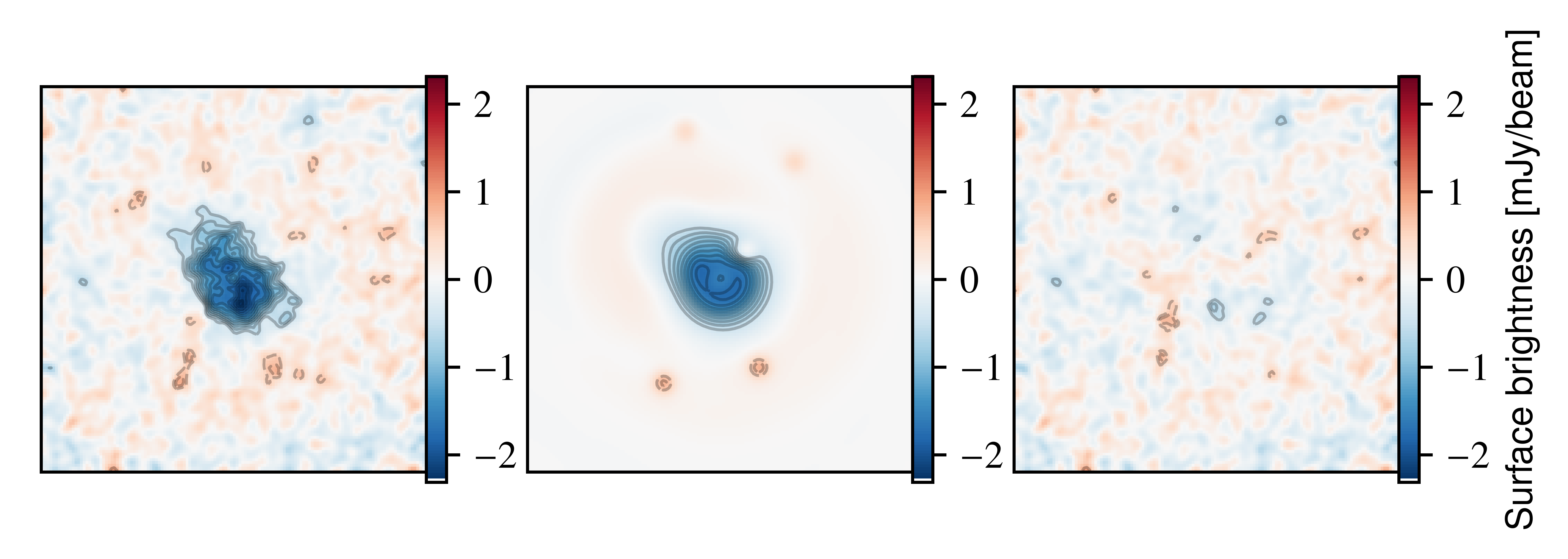}
\end{subfigure}
\caption{\footnotesize Single spherical model fit of the NIKA2 2 mm map for PSZ2G091 centered on the X-ray centroid coordinates (\textbf{top}), and 2-halo model (\textbf{bottom}). From \textbf{left} to \textbf{right}, we display the data, the model, with the point sources treated, and the residuals. The maps are given in a $5'\times5'$ area, and for display purposes, each map is smoothed with a gaussian kernel. The contours are showing the SNR level sets, starting at $3\sigma$ and increasing with a step of $1\sigma$.}
\label{fig:spherical_model}
\end{figure}

We first consider a single spherical halo centred on the X-ray centroid coordinates. A forward modelling approach is incorporated in an MCMC sampling framework to fit the parameters of the pressure profile, as well as the point sources, with the collaboration software {\tt PANCO2}\cite{2022EPJWC.25700024K}. We use a power law model, where each of the 6 bins follows the identity
\begin{equation}
\label{eq:power_law}
P(r)=P_i(r/r_i)^{-\alpha_i}.
\end{equation}
The top row of figure \ref{fig:spherical_model} shows the results of the fitting procedure. The spherical symmetry clearly does not encapsulate the bimodal nature of the cluster. It is then required to improve our modelling.

%
%
%
\noindent Then, instead of considering a single pressure profile, we jointly fit two halos at the positions of the X-ray peaks. %
The results of the fits are shown on the bottom row of figure \ref{fig:spherical_model}. It is clear that the two-halo model yields a more realistic representation of the dynamical state of the cluster. Additionally, the residuals are slightly improved in the region of the northern subhalo. Of course, due to the non-spherical nature of this cluster, it is not possible to consider a radial pressure profile. However, in section \ref{sec:thermo}, we describe how we recover an average radial profile for the two-halo model.
\section{Pressure profile reconstruction}
\label{sec:thermo}
Thermodynamical profiles are usually considered with the goal of reconstructing a mass profile, using the hydrostatic equilibrium assumption. This requires the presence of a 1D pressure profile, 
\begin{figure}[h]
\begin{center}
\includegraphics[scale=0.5]{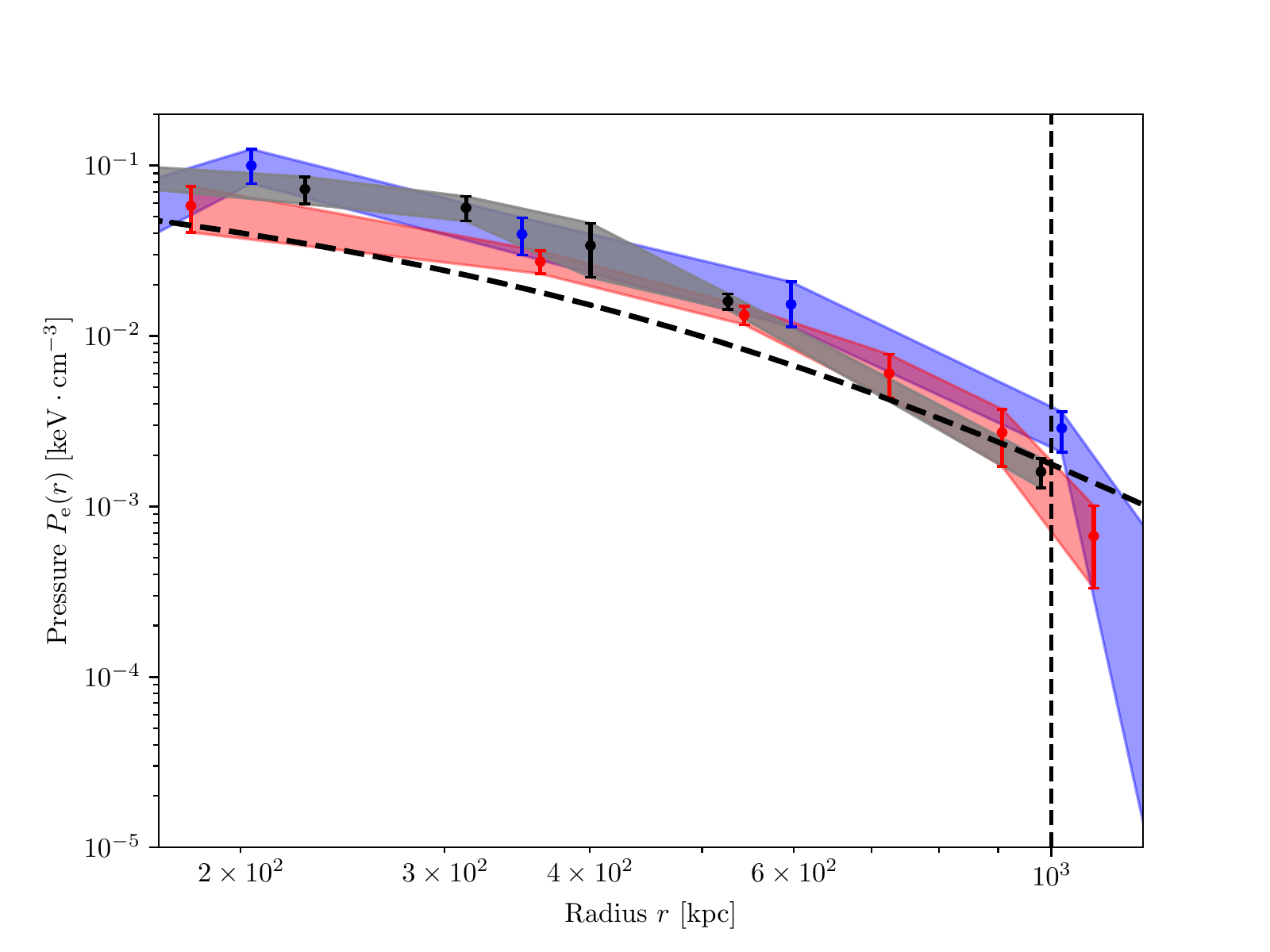}
\caption{\footnotesize In \textcolor{red}{red}, composite pressure profile obtained by combining the northern (N) and southern (S) profiles, as described in equation \ref{eq:mean_prof}. In \textcolor{blue}{blue}, profile obtained with the single spherical model. In black, we represent the contours obtained from XMM X-ray data only, for the spherical case.  The  black dashed line represents the universal profile. The vertical dashed line represents the radius $R_\mathrm{500}$ obtained from X-ray data assuming spherical symmetry.}
\label{fig:mean_pres}       
\end{center}
\end{figure}
when we previously fitted a 2D map. Thus, we use the following procedure to recover an average 1D profile. With the assumption of the two sub-halos lying in the same plane perpendicular to the line of sight, we recover a mean radial pressure profile by integrating both the profiles in annuli centered around the X-ray centroid coordinates. The pressure in the $i$-th bin reads:
\begin{equation}
\label{eq:mean_prof}
P_i=\left(\int_{r_i}^{r_{i+1}}(P_N(r)+P_S(r))2\pi r \mathrm{d}r\right)/\left(\pi(r_{i+1}^2-r_i^2)\right),
\end{equation}
where $P_i$ is the value of the pressure, $P_N$ and $P_S$ are respectively the profiles of the northern and southern subhalos, $r_i$ and $r_{i+1}$ being the inner and outer radii of the annulus. This allows us to reconstruct the profile shown in figure \ref{fig:mean_pres}, where we compare our reconstructed quantities with the universal profile \cite{2013A&A...550A.131P} . 
%
\section{Conclusions}
\label{sec:conc}
This analysis shows the challenges related to the complexity of cluster morphologies. In this work, we showed that for the case of the highly disturbed cluster PSZ2G091, taking into account the merging state of the cluster yields results that are slightly different from the spherical profile. This is a promising result, as the pressure profile impacts the $Y_\mathrm{500}-M$ relation. To complete this analysis, we plan to perform a full thermodynamical analysis of this cluster, recovering the 2D maps of physical quantities like the temperature and the entropy. This a precondition to assess the impact of the morphology of this cluster on its full mass reconstruction, and generally on cosmological inference using clusters.  

\section*{Acknowledgments}
\footnotesize{We would like to thank the IRAM staff for their support during the campaigns. The NIKA2 dilution cryostat has been designed and built at the Institut N\'eel. In particular, we acknowledge the crucial contribution of the Cryogenics Group, and in particular Gregory Garde, Henri Rodenas, Jean-Paul Leggeri, Philippe Camus. This work has been partially funded by the Foundation Nanoscience Grenoble and the LabEx FOCUS ANR-11-LABX-0013. This work is supported by the French National Research Agency under the contracts "MKIDS", "NIKA" and ANR-15-CE31-0017 and in the framework of the "Investissements d’avenir” program (ANR-15-IDEX-02). This work has benefited from the support of the European Research Council Advanced Grant ORISTARS under the European Union's Seventh Framework Programme (Grant Agreement no. 291294). E. A. acknowledges funding from the French Programme d’investissements d’avenir through the Enigmass Labex. A. R. acknowledges financial support from the Italian Ministry of University and Research - Project Proposal CIR01\_00010.}

\section*{References}
 \bibliography{references}

\end{document}